\documentclass[preprint,12pt]{elsarticle}

\usepackage{tikz}
\usetikzlibrary{shapes.geometric, arrows}

\usepackage{pgf-pie}

\usepackage{bchart}

\usepackage[breaklinks]{hyperref}

\usepackage{pgfplots}

\usepackage{pgfplots}

\usepackage{bchart}

\usepackage{graphicx}
\usepackage{amssymb}

\usepackage{lineno}

\usepackage{subfiles} 

\pgfplotsset{compat=1.16}

\begin{document}

\journal{Entertainment Computing}


\pgfplotsset{compat=1.16}

\begin{frontmatter}



\title{What Do We See: An Investigation Into the Representation of Disability in Video Games}
 \tnotetext[label1]{}
\author{Dr. Jethro Shell}
\ead{jethros@dmu.ac.uk}
\address{Gateway House, The Gateway, De Montfort University, Leicester, LE19BH, United Kingdom\fnref{label3}}




\begin{abstract}
There has been a large body of research focused on the representation of gender in video games \cite{williams2009virtual}\cite{ivory2006still}\cite{near2013selling}. Disproportionately, there has been very little research in respect to the representation of disability. 
This research was aimed at examining the representation of disabled characters through a method of content analysis of trailers combined with a survey of video gamers.
The overall results showed that disabled characters were under-represented in video games trailers, and respondents to the survey viewed disabled characters as the least represented group.
Overall, both methods of research concluded that the representation of disabled characters was low. Additionally, the characters represented were predominantly secondary, non-playable characters not primary. However, the research found that the defined character type was a mixture of protagonists and antagonists, bucking the standard view of disabled characters in video games.

\end{abstract}


\end{frontmatter}



\section{Introduction}
\label{S:1}
There has been a number of pieces of research that have focused on the character depiction in video games (video games and computer games will be used interchangeably within this document). The predominant focus of this has been on gender with a smaller sample of discussions involving race and ethnicity. This piece of research has set out to investigate an area that has received far less attention, namely disabled representation of characters in video games. This research investigated how the video games industry has addressed disabled characters by analysing the quantity of characters that can be defined as being disabled and the nature of their portrayal over a ten year period ($2006$ - $2016$).
To discuss the representation of disabled characters in video games, there first needs to be the adoption of a definition of disability. It is challenging to define as it has a wide ranging scope. The UK government in the Equality Act of 2010 define disability as \begin{quote}``a physical or mental impairment that has a ``substantial" and ``long-term" negative effect on your ability to do normal daily activities"\cite{disabilityact2010govuk}.\end{quote} 
Substantial is classed as being more than minor or trivial, for example impacting getting dressed. Long term is defined as over 12 months in length including the development of conditions \cite{disabilityact2010govuk}. Under UK law there are nine characteristics that are protected from discrimination, disability is included in this list. The definition used in this context is the same as the definition used in the Equality Act of 2010. Within this there is an expansion on this definition. The guidance goes on to say:
\begin{itemize}
\item Mental impairment may include mental health conditions e.g depression, learning difficulties e.g dyslexia and learning disabilities.
\item Severe disfigurement is protected as a disability without having to demonstrate the same adverse effects as other conditions. This does not include examples such as tattoos and piercings for non-medical or decorative purposes \cite{legaldefinition2013jisc}.     
\end{itemize}
This research has used the definition described as a reference for the content analysis.

The focus on disability representation can find resonance in the analysis of video games and gender representations, and the representation of ethnicity and minority groups in video games. In Sections \ref{State:Gender2.1} and \ref{Ethnic:2.1}, a short review will be carried out looking at the current analysis of representation of gender and ethnicity in video games.

\section{State of the Art}
\label{State:2}

\subsection{Gender Representation in Video Games}
\label{State:Gender2.1}

A number of studies have looked at the variation in the portrayal of genders in video games. Content analysis has been the prime form of research. Predominantly it found that women were absent from a large portion of the video games analysed and those included were depicted in a more sexualised manner than their male counter-parts \cite{behm2009effects}.
Dietz \cite{dietz1998examination} carried out work in this area. They looked at a sample of $33$ popular Nintendo and Sega Genesis video games. It was found that there were no female characters in $41$\% of the games where characters were included. In the subsequent games, $28$\% of women were portrayed as sex objects and only $15$\% were viewed as heroes or as action characters. Dietz also noted that of the action characters, the clothing was in keeping with stereotypes of females, e.g females fighters in Mortal Combat II wore ``high boots, gloves and revealing leotards" \cite{dietz1998examination}. It was also noted that of the sample, $21$\% were presented as the victim, again stereotyping the gender. It should be noted that the content analysis taken looked solely at Nintendo and Sega games and was carried out in the Spring of 1995.
Glaubke et al.\cite{glaubkefair} also looked at the differences in prominence of players in video games. $70$ games were assessed in total, with 1716 characters. Of those games, $64$\% contained male characters in total. Of the player controlled characters, only $12$\% were female. Half of the female characters were subsidiary or "props", while male characters were defined as being either the protagonist or the antagonist. Similarly to the evidence proposed by Dietz, stereotypes were prevalent in the games. Male characters were more likely to engage in physical aggression ($52$\%), where as females were four times more likely to be nurturing ($8$\% compared to $2$\%). Overall, the research of Glaubke substantiated many of the aspects that were found by Dietz's study, that being female characters were less prominent and had smaller roles compared to their male counterparts.
Lynch et al\cite{lynch2016sexy} further commented that sexualisation of female characters had an effect on women's perceptions of video games. Whilst substantiating some of the discussions that Lynch et al. put forward, Hartmann and Klimmit\cite{hartmann2006gender} found through evidence based research ``that due to the emergence of female subcultures adopting contemporary video games designed for males, and the advent of new games that successfully engage female players, the gender gap has started to narrow" \cite{hartmann2006gender}.

\subsection{Race and Ethnicity in Video Games}
\label{Ethnic:2.1}
In comparison to the analysis of gender, ethnicity and race have had far less focus. Mou and Peng\cite{mou2009gender} comment that a reason for this may be a result of a conscious avoidance by game makers of specific race representations. 
With that said, there has been some work. Passmore and Mandryk\cite{passmore2018gaming} produced a qualitative survey to examine the concept of race and ethnicity as perceived by ethnically diverse video gamers. They were driven by a view that "players of colour desire greater diversity in game characters and wish to "self-represent in games"\cite{passmore2018gaming}. This could come about through the use of a greater number of customisations for players. Gardner and Tanenbaum \cite{gardner_tanenbaum_2018} agreed with this,  finding that it is difficult to produce a consensus of the race and ethnicity of non-white players. The difficulty, comes from the need to represent the subtlety of the characters. They commented that presentation can be missed producing technical challenges in the creation of areas such as African-American hair and culturally accurately gait\cite{passmore2018gaming}, for instance.  

Parallels can be drawn between disability representation with both gender and race. Although Section \ref{State:2} discusses some of the research in this area, further work is needed but this is beyond the scope of this research. 
In the following section, Section \ref{Methodology:1.1}, the process used to gather information for the content analysis will be given. 
\section{Methodology}
\label{Methodology:1.1}
The methodology used for the content analysis of the video game trailers is based on three sections;

\begin{enumerate}  
\item Scraping of data from key on-line review sites to highlight prominent games. 
\item Cleaning and processing of the data to define a list of 108 (two games were not available from 2016) that represent a selection from each of the years between 2006 and 2016. The data was defined as being composed of:
	\begin{enumerate}
	\item the largest on-site reviews.
	\item the highest user score.
	\item having at least $10$ or higher users that had submitted a review ($10$ users were defined as the value that reduced the impact of low numbers of user reviews whilst maintaining the use of games that had comparatively large site reviews). 
	\end{enumerate}
\item Examination of trailers of each video game to extract content to identify any main protagonist or antagonist that are represented as being disabled. 
\end{enumerate}
This methodology follows a similar approach to that of Ivory \cite{ivory2006still}. They focused on the use of "online video game reviews as an indirect measure of video game content". The focus of this work was on gender representation in video games, a subject that is covered in Section \ref{State:2}.
\subsection{Data Set and Data Acquisition}
As was highlighted previously, the focus of this research was on the changing landscape of games over $10$ years between 2006 and 2016. To these ends, the acquisition of data was focused on this time period. The overall data set was compiled using a combination of publicly available data. Data was acquired from three prominent game review sites, Metacritic.com, VGChartz.com and IGN.com. The VGChartz data was produced by Smith\cite{smith_2016} which expanded on data processed by Kirubi\cite{kirubi2016} adding data from the Metacritic website. The final addition was a scrape of data by Grinstein \cite{grinstein_2016}) which was taken from the IGN website. This provided review scores from both the sites themselves and users. The combined data set is available at Kaggle\cite{kaggle_2019}. 
\subsection{Data Set Processing}
\label{sec:dataset}
The data acquired went through a process of cleansing. Initially the focus was on the IGN data. Using the Metacritic data set as a base, all reviews that were on mobile platforms were removed, for example all games for the iPhone. This was to allow the focus of the reviews to be on console and personal computer (PC) formats. All reviews earlier than 2006 were removed from the data set alongside any game that had no reviews at all. As with the IGN data set, the Metacritic data similarly had any game without a user or review score removed. It is believed that this provides a more robust data set in which to examine the proposed research question. It reduces the impact of games that have little interaction with the review and sales process, whilst having large user score values. The data set produced by Grinstein \cite{grinstein_2016} was also cleansed in a similar way. This produced the final data set that was used. 
A weighted value was used to construct the final 108 game set. The value was defined as:
\begin{equation}
i = \frac{((j / k) + l + m + n + \frac{p - \min{p}}{\max{p} - \min{p}})}{q}
\end{equation}
were \textit{i} is the weighted value, \textit{j} is the score given by Metacritic critics, \textit{k} is the divisible value to return a score between percentage, \textit{l} is the Metacritic user critic score, \textit{m} is the IGN user score, \textit{p} is a normalised value of the global sales for a video game as supplied by VGChartz and \textit{q} is a value to convert the weighting into a percentage.
This value is believed to represent a view across the data supplied and is in line with a number of other metrics. This can be seen in the five descriptors of games defined within IGN's phrase variable. The variable 'masterpiece', the highest level, was within $27$ of the $108$ games.
As the focus of this research is to investigate the representation of disabled characters in video games, $10$ of the highest valued games for each year based on the weighted value were taken as the data set to be used. It is believed that this represents the most prominent games over the ten year period. Other studies have taken a similar approach in not using a random selection of video games\cite{dietz1998examination}.
\subsection{Character Examination}
Each of the games within the data set were examined by the process outlined in Section \ref{sec:dataset}. A modification of the methodology first defined by Ivory \cite{ivory2006still} for studying gender in video games was used. Ivory's method used on-line reviews as an indication of how prominent the games were within the gaming world. The method was altered here by using trailers rather than reviews. This is more in keeping with the video game industries use of trailers as a form of information, marketing and pre-sales.
The use of trailers to assess the representation of the characters was also proposed by Mou and Peng\cite{mou2009gender}. They used an approach to identify whether characters were male or female. A character was defined as being a human, animal or an object that has human-like appearance or qualities. The nature of this study refines this characterisation as focusing solely on human and human form characters (the latter being those that may encompass the fantasy genre). This analysis took a similar approach looking at characters in the game that were of human form. In Downs and Smith's \cite{downs2010keeping} analysis of gender in games, they focused on primary and secondary characters. These were defined as;
\begin{enumerate}
	\item Primary characters were the protagonist that a game player could actively manipulate which the success or failure of the game revolved around.
	\item Secondary characters were immediately tied to the game play but were predominately non-playable.
	\item Tertiary characters were those that assisted or helped to make up the environment (people in crowds of on-lookers, monsters in hoards that form set pieces).
\end{enumerate}
As with Downs and Smith's \cite{downs2010keeping} study, the focus here was on primary and secondary characters with the inclusion of tertiary characters in the second group.
\section{Game Data Set Analysis}
\label{S:5}
\subsection{Disabled Characters Within Game Trailers}
\label{subsec:CharactersInTrailers}
The data produced from the cleaning process created a set of $108$ reviewed trailers in total. After examination of these, $20$ of the trailers ($18.52$\%) were deemed to have a disabled character shown. Of these, $15$ were main characters with $10$ being playable. Placing all of this analysis together, the trailers showed a total representation of a playable disabled characters as $0.99$\%. However, of the disabled characters included, $75$\% were primary or secondary characters that played an influential role. This is an indication that despite the low inclusion of disabled characters, they are integral to the story and often drive it forward where included. 

A prime example of this is the character of Lester Crest in Grand Theft Auto V (GTA V), the highest weighted value game in the data set. Lester Crest is described as suffering from a wasting disease which has limited his mobility. Due to this, he has also become overweight and has asthma\cite{lesterCrest2019}. Lester Crest is affiliated with Michael De Santo, a main playable antagonist of the game. Lester interacts substantially in GTA V's story. He is a regular member of heist crews, working mostly as a planner and information scout. This does, however, show him in a negative light although GTA V does not have a clearly defined "good vs bad" element.

Of the 20 trailers including a disabled character, a greater number featured them as being protagonists rather than antagonist characters. This opposes many of the stereotyped views of disabled characters as being the "villain". 
It can be said that this relates to the representation of disabled characters having a form of augmentation to enhance their powers. The analysis clearly demonstrated how this conclusion can be drawn. The survey highlighted a  number of characters that gained technology or special powers to overcome their disability. For example, in the game MadWorld Jack Cayman\cite{jackcayman_2020}, the main playable character has a chainsaw attached to a prosthetic arm. In Metal Gear Solid 4 the main antagonist, Solid Snake, has developed a degenerative condition due to the cloning method used to create him. He requires the use of a nano-tech injection to maintain his health. However, Solid Snake is not shown to be enhanced by the technology and its interaction with his condition but that the technology is used just to maintain his existence\cite{solidsnake_2020}.
The games above (a small sample) demonstrate the use of "additions" or "enhancements" to allow for the character to exist or function in the game world. This shows an additional section that the depiction of disabled characters fall into, that of the "tech can overcome" trope. The author suggests that, as previously discussed, the disability of a character drives part or all of the narrative, and can not be a part of the narrative or virtual world without this. It can always be argued that as a fictional world, the backstory of the characters is of interest and the physical representation of this is an outcome.  

\subsubsection{Examination of Characters By the Defined Dates.}
\label{subsubsec:DefinedDate} 
Over the analysed period (2006 to 2016), there were no discernible increases or decreases in the numbers of disabled characters that were portrayed. Table \ref{tab:year} displays these values. 

\begin{table}[ht]
\centering
\resizebox{\textwidth}{!}{%
\begin{tabular}{|c|c|c|c|c|l|l|l|l|l|l|}
\hline
\multicolumn{11}{|c|}{Year Of Release}                                     \\ \hline
2006 & 2007 & 2008 & 2009 & 2010 & 2011 & 2012 & 2013 & 2014 & 2015 & 2016 \\ \hline
1    & 0    & 1    & 3    & 3    & 1    & 0    & 2    & 2    & 1    & 1    \\ \hline
\end{tabular}%
}
\caption{Number of Disabled Characters.}
\label{tab:year}
\end{table}

The survey carried out in Section \ref{sec:6} supports this as players described the representation of disabled characters as being low. Based on this and the findings of the survey, it can be inferred that little change has been made in terms of the inclusion of disabled characters.
\subsubsection{Examination of Characters By Sales.}
\label{subsubsec:Sales}
 The global sales of games and its relationship to disabled representation was also analysed. Looking at the video games that included a disabled character, the sales were higher than those without. However, the games in that group were from franchises that contained some of the highest selling across the board. For example, Grand Theft Auto V sold $21.04$ million copies up until $2016$ with its predecessor, Grand Theft Auto IV selling $11.01$ million copies. Overall the games from this franchise up until 2016 had sold $38.4$ million copies. As previously discussed, the Grand Theft Auto series of games contain the character, Lester Crest, so possibly skewing the output. Similarly the Metal Gear Solid franchise, which has a playable main protagonist with a disability, has a global sales figure of $10.98$ million.
This is not to say that singularly there are games that do not contain disabled characters that compare in terms of sales. The FIFA games are one of the highest selling franchises and continue to be so as new versions of the game are released each year\cite{FIFA2019}.
\subsection{Discussion}
\label{subsection:quandis}
Compiling together all of the findings of the analysis of the trailers, it is clear that a number of elements are apparent.
\begin{enumerate}
\item{Analysis of the trailers showed a low depiction of disabled characters ($18.52$\%).} 
\item{Within the trailers assessed, only $0.99$\% showed a depiction of a playable disabled character.}
\item{The appearance of the disabled characters was predominantly of an protagonist rather than a antagonist.}
\end{enumerate}
It can be suggested that the lack of disabled characters can be attributed to two issues within games development:

\begin{enumerate}
	\item Inability of developers to remove themselves from the stereotype trap: Games developers, specifically those involved in the formation of the narrative, find it difficult to extract themselves from stereotypes. When looking at gay, lesbian, bisexual, and transgender (GLBT) representation in games, Shaw\cite{shaw2009putting} discussed that there were few games that introduced complex sexual relationships. Instead a stereotyped default "norm" of a standard heterosexual relationship was used. Disability representation can be said to follow the same track. The stereotyped view of a hero, for instance, is of a male, highly muscular figure often with unrealistic proportions. The extraction of the representation of disability from the stereotypical representation proves to be problematic. 
	Sylvia et.al \cite{sylvia2014virtual} discuss that "for men regularly exposed to images of unrealistic male body ideals, the unrealistic body could eventually become the benchmark for a normative body". Taking this in mind, the continuation of the games industry to focus on the hyper-masculine or overtly sexualised female (although it can be said some small steps have been taken in this area) protagonist imposes difficulties in the ability to define a disabled individual as a main playable character. Instead they fall into the stereotyped trap of defining those with disabilities as a secondary antagonist.
	\item Narrative based around the representation of a character: As with other media, the representation of a character in a video game more often than not drives the narrative. This is to be expected. Those groups that have a lower representation in games appear to be used in this way more predominately. Women and the BAME (Black, Asian and Mixed Ethnicity) community are key examples of this. For example, games such as Mafia III partially drive the story through the background of the main protagonist Lincoln Clay. Clay is believed to have an Italian father coupled with Afro-Caribbean heritage. The story takes place in Southern America at a time of extreme racism. The narrative of the story is driven by the nature of the characters representation as a person of mixed ethnicity\cite{plante_2016}. 
\end{enumerate}
\section{Qualitative Analysis of Game Representation}
\label{sec:6}
In this section the representation of disability in video games is analysed through the use of a qualitative survey.  
The survey was carried out using a combination of both closed and open questions. This allowed participants to answer in an defined manner, but also expanding where necessary. There were the use of two point questions taking the form of Yes / No, multi-point questions for example, Yes / No / Prefer not to say and open questions. 
The survey took place over a period of $25$ days. All of the participants had at one stage played a video game, although this was not a requirement of the survey. Over this period, $76$ responses were gathered. In the subsequent sections, a discussion will take place analysing the responses.
\subsection{Analysis of the Representation Survey}
\label{sec:6.1}
The respondents of the survey fell into a number of categories. When asked their age, the majority returned a value of between $18$ - $25$ years. The second largest group was $26$ - $35$ year old's at $32.89$\% of the total. $18$ - $35$ years being the largest group follows the trend of other surveys. In 2019 a survey in the United States of America (USA) showed that $40$\% of players had an average age of $18$ - $35$ years \cite{statista_age_2019}. 
The breakdown of the age supplied can be seen in Figure \ref{fig:age}.

\begin{center}
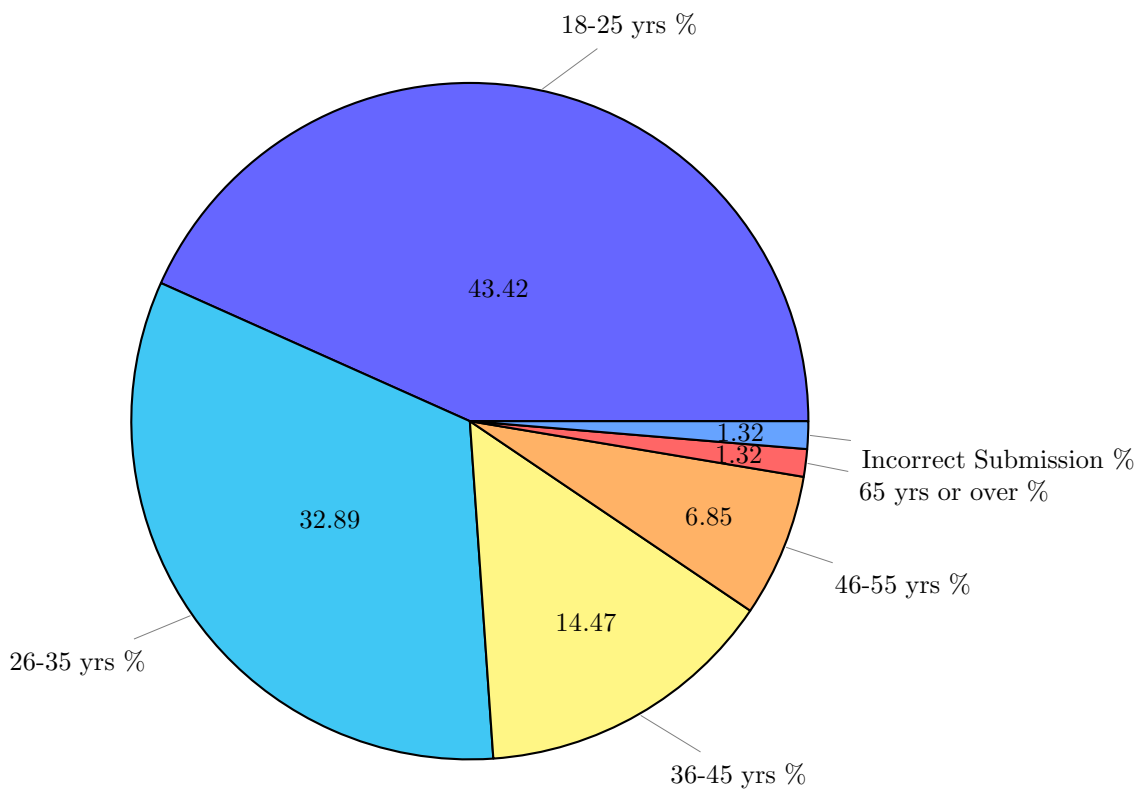
\begin{figure}[ht]
\begin{tikzpicture}
[scale=1.5]
\tikzstyle{every node}=[font=\footnotesize,every only number node/.style={text=black}]
\pie [sum=auto,after number=,text = pin]{43.42/18-25 yrs \%, 32.89/26-35 yrs \%, 14.47/36-45 yrs \%, 6.85/46-55 yrs \%,1.32/65 yrs or over \%,1.32/Incorrect Submission \%} 
\end{tikzpicture},
\caption{Respondents Identified Age Groups} 
\label{fig:age}
\end{figure}
\end{center}

Participants were also asked in which decade they began to play video games. The largest group cited the 1990's (31.58\%), closely followed by the 2000's. This links to the emergence of the Millennial generation (those born between 1981 and 1996 \cite{dimock_dimock_2019}) who are poised to overtake the previous generation, Generation X, in terms of global spending power \cite{tilford_2018}. Spare income allows an increase in the purchase of non-essential items such as entertainment. Additionally, as the video games industry has matured, this generation have engaged more than previous generations. 

The survey also asked which gender they identified with. The groups were:
\begin{itemize}
	\item Male.
	\item Female.
	\item Non-Conformist.
	\item Non-Binary.
	\item Transgender Male.
	\item Transgender Female.
	\item Prefer Not To Say.
	\item Other
	\item Add Option
	\end{itemize}
The ratio of male to female participants were more in favour of males. In this survey 76.32\% of the respondents were male compared to only 14.47\%. This bucks the trend seen in other surveys that shows that the male / female ratio of gamers is far closer to equal. Over the past 10 years female gamers have moved from being $\frac{2}{5}$ of the gaming population to just under equal (46\% of gamers \cite{statista_age_2019}). This indicates that this survey may fall outside of the standard sample and further surveys may need to be carried out to demonstrate a more representative group. The survey also included members of the non-conformist and non-binary groups. There is little evidence, and it is outside of the scope of this journal, that these groups have been investigated in terms of representation in video game surveys. The full break down of the results can be seen in Figure \ref{fig:gender}.

\begin{center}
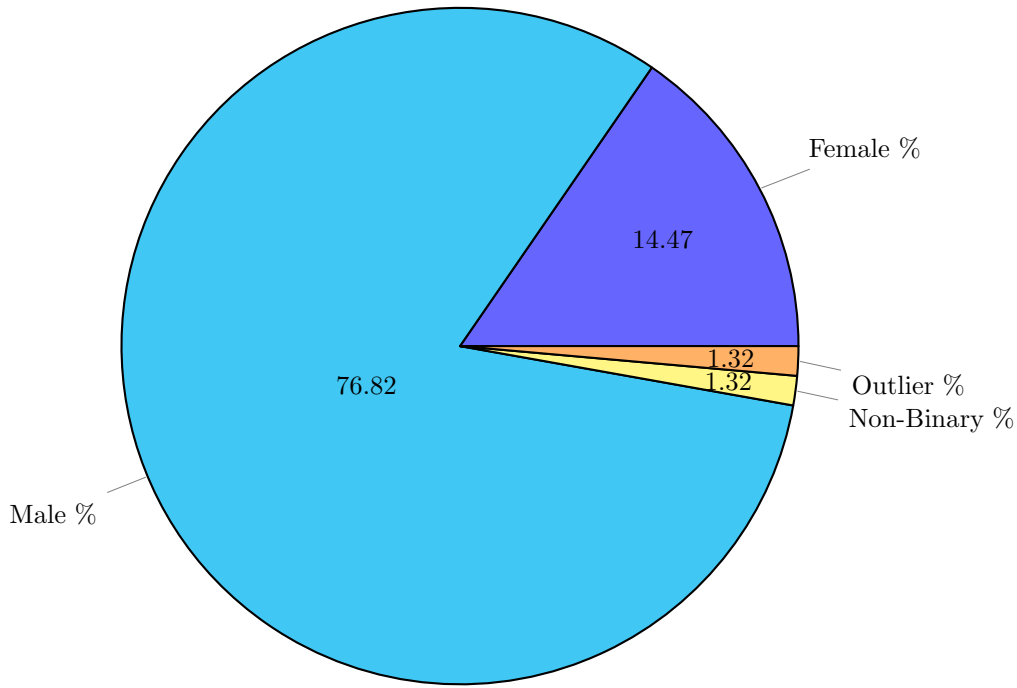
\begin{figure}[ht]
\begin{tikzpicture}
[scale=1.5]
\tikzstyle{every node}=[font=\footnotesize,every only number node/.style={text=black}]
\pie [sum=auto,after number=,text = pin]{14.47/Female \%, 76.82/Male \%,1.32/Non-Binary \%,1.32/Outlier \%} 
\end{tikzpicture},
\caption{Respondents Identified Age Groups} 
\label{fig:gender}
\end{figure}
\end{center}

In order to gauge whether having a disability impacts how video games are participated in and viewed, participants were asked if they considered themselves to have a disability. 35.53\% said "Yes", 57.89\% said "No" and 6.58\% preferred not to say. Of those that defined themselves as having a disability, the most prominent was those with Autism Spectrum Disorder (ASD). It should be noted that ASD was listed alongside other conditions such as Dyslexia. ASD lacks a unifying definition, however, Goldstein and Oznoff\cite{goldstein2018assessment} comment that a "conceptualisation of ASD is a biologically determined set of behaviours that occurs with varying presentation and severity, probably as a result of varying causes". Rutter\cite{rutter1983cognitive} found that patterns of behaviour of individuals with ASD were distinct from others with intellectual disabilities. They found that individuals had (to varying degrees) the possibility of issues with;
\begin{itemize}
	\item Language and language related skills.
	\item Semantics and pragmatics.
	\item Perception.
	\item Memory weakness.
	\item Cognitive problems.
	\item Impairment in social relations.
\end{itemize}
Within the survey, 50\% of those that defined themselves as having ASD rated their game play as being "Very frequently" or "Somewhat frequently". "Somewhat frequently" and "Very frequently" ranged from 1-2 hours to 5-7. Those with other defined disabilities (7 of 27) also had high ratings of play, incorporating "Somewhat frequently" and "Very frequently", however the values of the average play per day were lower (three players at 3-4 hours, two at 1-2 hours and a single respondent at less than 1 hour).
\pgfplotstableread[row sep=\\,col sep=&]{
    interval & ASD & Other  \\
    less than 1hr     & 38.46  & 7.69  \\
    1--2hrs	& 15.38 & 23.08    \\
    3--4hrs     & 30.77 & 30.77    \\
    5--7hrs   & 7.69 &  7.69  \\
    Undefined & 7.69 & 30.77  \\
    }\mydata  
\begin{center}
\begin{figure}[ht]
\begin{tikzpicture}
    \begin{axis}[
            ybar,
            bar width=.5cm,
            width=\textwidth,
            height=.5\textwidth,every node near coord/.append style={font=\tiny},
            legend style={at={(0.5,1)},
                anchor=north,legend columns=-1},
            symbolic x coords={less than 1hr,1--2hrs,3--4hrs,5--7hrs, Undefined},
            xtick=data,
            nodes near coords,
            nodes near coords align={vertical},
            ymin=0,ymax=100,
            ylabel={\% of group},
            every node near coord/.append style={font=\tiny}
        ]
        \addplot table[x=interval,y=ASD]{\mydata};
        \addplot table[x=interval,y=Other]{\mydata};
        \legend{ASD, Other}
    \end{axis}
\end{tikzpicture}
\caption{Hours of Play Across Groups.}
\end{figure}
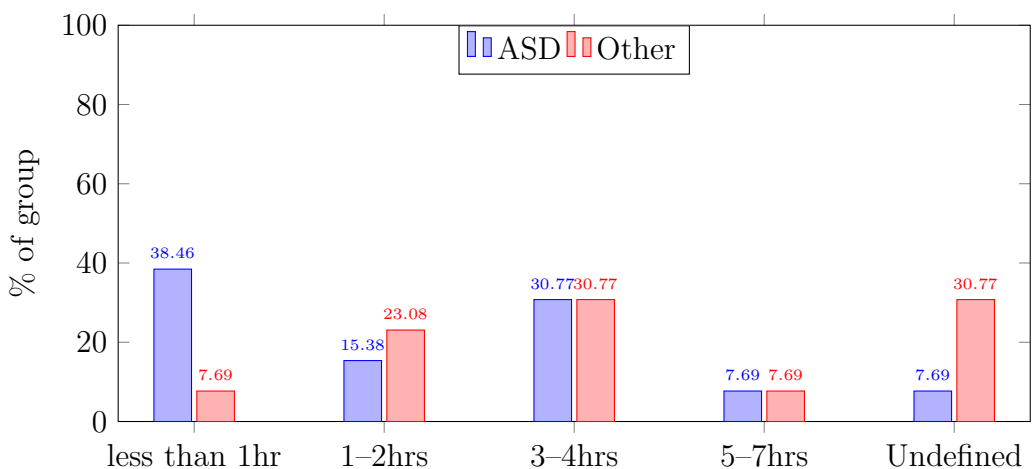
\end{center}

Of the group that defined themselves as having ASD, the hours that were played was a smaller quantity than none ASD individuals. This is opposed to the view that those with ASD are prone to problematic video game use\cite{engelhardt2014video}. Research by Mazurek et al. found that boys with ASD spent more time than boys with traditional development (2.1 vs 1.2 hours per day)\cite{engelhardt2014video} playing games. 
Whilst the research by Mazurek et al. should be be taken into account, the survey used covered differing gender definitions which may have altered the scope of the results. 
To understand the perspective of those surveyed in regard to changes in character portrayal and the manifestation of this, a number of questions were posed. A general question was asked, "Do you feel that video games genres have changed since you started to play?"

\begin{center}
  \begin{tabular}{ l|l|l|l}
    \hline
    & Disabled & Non-disabled & Prefer Not To Say \
    \\ \hline
    Not At All & 0\% & 9.09\% & 20.0\% \\
    Not So Much & 0\% & 36.36\% & 20.0\% \\
    Slightly & 14.81\%  & 34.09\% & 0.0\% \\
    Moderately & 25.92\% & 31.81\% &60.0\% \\
    Very Much So & 59.26\% & 54.54\% &0.0\% \\
    \hline
  \end{tabular}
\end{center}
The results showed little difference in views between the groups at the higher end, however the lower values (for example "Not At All" and "Not So Much") showed a greater difference. 
The respondents were given an open question that focussed on the areas of change. The most referenced group was gender with 54 of all participants citing this character trait. This was closely followed by sexuality and ethnicity. The responses showed that players are aware of changes in characters depiction, although there were mixed opinions about what result it had caused. One statement made commented that:
\begin{quote}"It's more sexual and it's more detailed but I wouldn't say it is more progressive in terms of minorities. \end{quote}
The sexualisation of characters were often referred to. This demonstrates again that players are aware of the changes in the representation of different genders in video games but that these changes are not always perceived as being a move forward.

The most absent group was disability. For disabled characters, 12 of the overall respondents made reference to them. This is amplified by the a disparity between the disabled and non-disabled groups. There were a higher number of disabled players that viewed a difference (29.62\% to 9.09\%). When combining the response to the two questions, the non-disabled players defined that video games genres had changed slightly. The disabled respondents viewed the changes as being higher, ranging from "Moderately" at the lowest point to "Extremely". Where disability was not included and the respondents were disabled, the level of change was far lower including "Not At All". 

The disabled players recorded less of a change in genre when disability was not included. Non-disabled players recorded a higher change, however it needs to be noted that the group samples were small. There is an indication from these results that the disabled group saw a lack of change in genres that included disability. 

The greatest difference between the responses of the disabled and non-disabled groups were in playing as a character with disabled traits. The break down of values can be seen in the Figure \ref{fig:DisabledCharacters} and Figure \ref{fig:DisabledCharacters2}.
\begin{center}
\begin{figure}[ht]
\begin{tikzpicture}
\pie [rotate = 180,text = legend]
    {11.11/No, 7.40/Only Played As a Character That Is Predefined, 48.14/Sometimes, 33.33/Yes}
\end{tikzpicture}
\caption{Disabled Players That Have Played As a Disabled Character.} \label{fig:DisabledCharacters}
\end{figure}
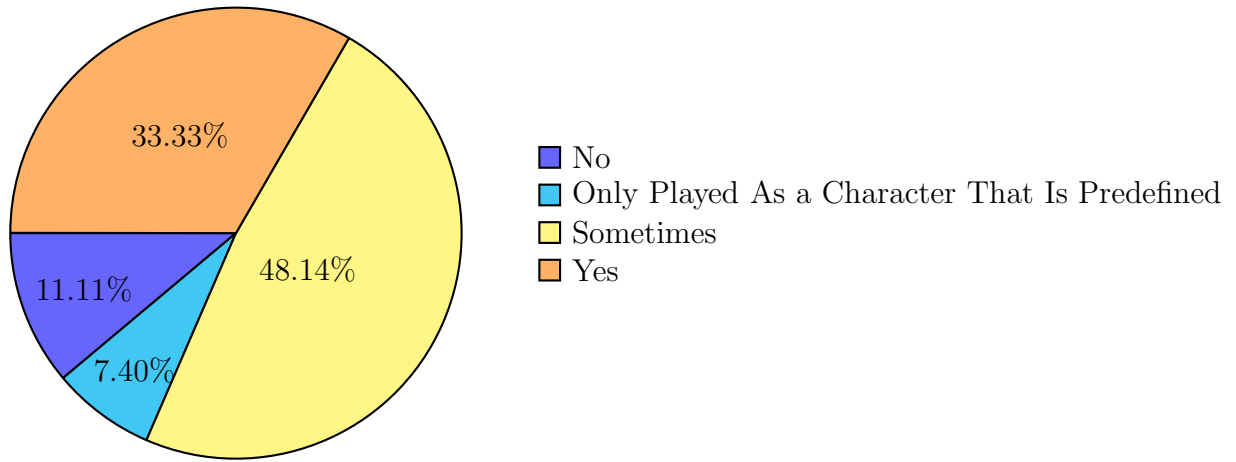
\end{center}

\label{Confused Question}
\begin{center}
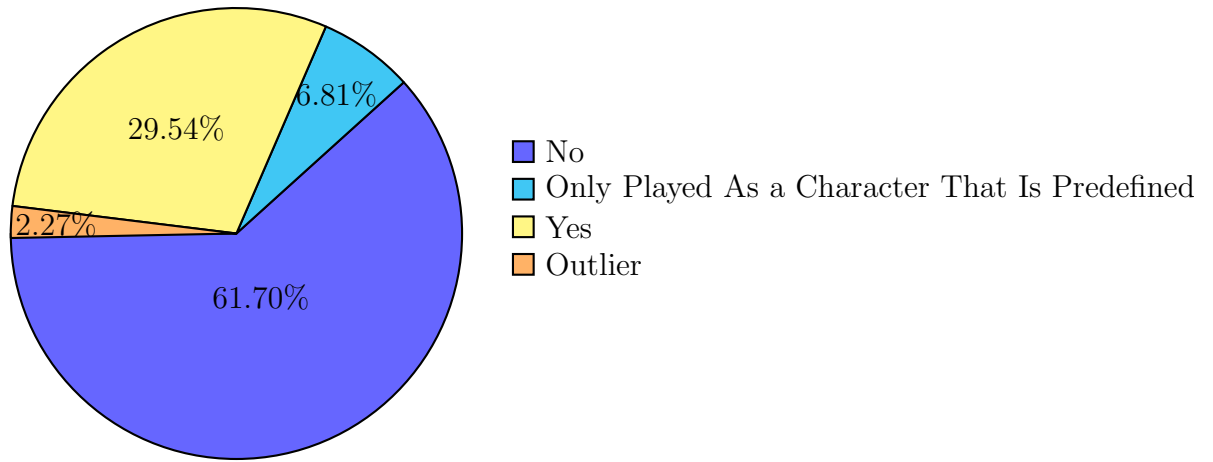
\begin{figure}
\begin{tikzpicture}
\pie [rotate = 180,text = legend]
    {61.70/No, 6.81/Only Played As a Character That Is Predefined,  29.54/Yes, 2.27/Outlier }
\end{tikzpicture}
\caption{Non-disabled Players That Have Played As a Disabled Character.} \label{fig:DisabledCharacters2}
\end{figure}
\end{center}
It is clear from this, and the responses to the previous question, that there is a strong relationship between a disabled player and an awareness of disabled characters. The fact that disabled gamers play more as disabled characters could be attributed to their understanding of how certain elements of a disability may affect an individual. This may be especially true of those that strive to have their own disability represented, much in a similar way to both gender and ethnicity.
There were also differences in the consideration of how applicable the disabled character was to the game. A breakdown of the numbers can be seen in Figure \ref{fig:disabledCharacter}

\pgfplotstableread[row sep=\\,col sep=&]{
    interval & Disabled & Non-Disabled & Prefer Not To Say  \\
    No     & 11.1  & 0.0  & 50.0\\
    No: Some Degree	& 0.0 & 7.7 & 0.0  \\
    Yes: Some Degree    & 44.4& 53.8   & 50.0 \\
    Yes   & 55.5 &  38.5 & 0.0 \\
    }\mydata 
\begin{center}
\begin{figure}
\begin{tikzpicture}
    \begin{axis}[
            ybar,
            bar width=.5cm,
            width=1\textwidth,every node near coord/.append style={font=\tiny},
            height=.5\textwidth,every node near coord/.append style={font=	                    \tiny},
            legend style={at={(0.5,1)},
                anchor=north,legend columns=-1},
            symbolic x coords={No,No: Some Degree,Yes: Some Degree,Yes},
            xtick=data,
            nodes near coords,
            nodes near coords align={vertical},
            ymin=0,ymax=100,
            ylabel={\% of group},
            every node near coord/.append style={font=\tiny},
        ]
        \addplot table[x=interval,y=Disabled]{\mydata};
        \addplot table[x=interval,y=Non-Disabled]{\mydata};
        \addplot table[x=interval,y=Prefer Not To Say]{\mydata};
        \legend{Disabled, Non-Disabled,Prefer Not To Say}
    \end{axis}    
\end{tikzpicture}
\caption{Applicability of Characters\label{fig:disabledCharacter}}

\end{figure}
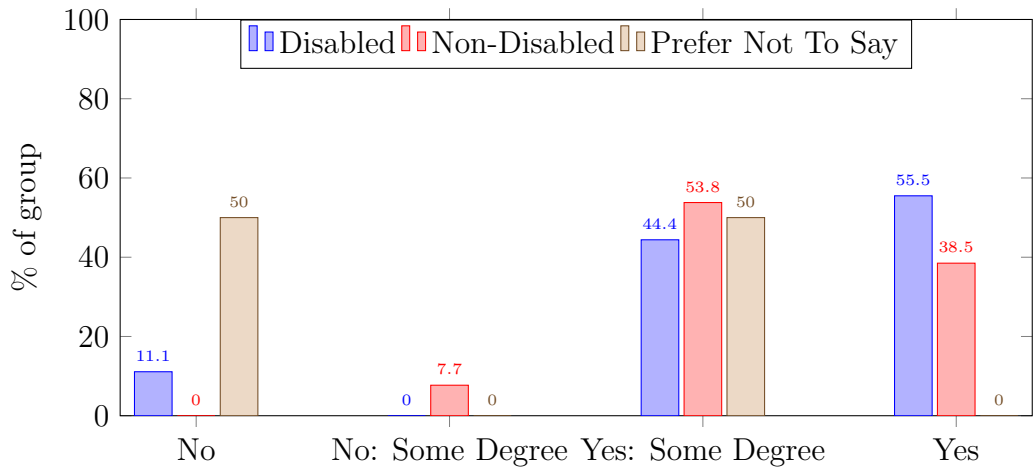
\end{center}
The disabled group felt that the characters portrayed were overall more applicable to the game compared to the other groups. To understand these responses further, this was followed by an open question. Those that responded negatively were asked to describe the reason why. Both the disabled and the non-disabled groups largely responded that they had never played a game with a disabled character. This was surprising and appeared to represent a misunderstanding of the previous question. It is, however, interesting for both sets of players to comment they have never played as a disabled character indicating some lack of the inclusion of disabled characters. This comment must, however, be taken against the confused nature of the responses.

Responses to the open question differed based on the contrasting answers to the previous. One of the "Yes" responses commented that: \begin{quote} "Yes, but it is usually dramatised to a degree since the main purpose of a game is for entertainment, so no can be argued."\end{quote} The purpose of video gaming and its representation of the real world is a complex question to answer. Any game world that sits outside of everyday life such as a post-apocalyptic game world, for example The Last Of Us\cite{LastOfUs2019} or the Fallout series\cite{Fallout2019}, or a fantasy universe such as the Witcher games\cite{Witcher2019} or World Of Warcraft\cite{WOW2019} can be argued to not be overall representative, so the inclusion of different characters are more concerned with the telling of a strong story rather than demonstrating the diversity of a population. The previous quote does, however, hint towards the fact that video game producers use disability to drive story rather than as a representative tool. Another individual agreed with this, commenting that the inclusion of disabled characters could be a plot device which can be "forced or offensive".
This confirms, in part, the findings of the trailer analysis. Stereo-typically, prosthetic or enhancements are often shown to overcome disabilities whilst enhancing elements of the character. This can be a plot device or as the main narrative of the game. Prime examples of this can be seen in:
\begin{itemize}
\item{Barret Wallace from Final Fantasy VII who has a Gatling gun replacement for an arm that was destroyed by security forces.\cite{robison_2011}}.
\item{Wolf from Sekiro: Shadows Die Twice initially has both arms. Following a battle with the first boss of the video game, he loses his left arm which is then replaced with an arm that is equipped with a number of sophisticated elements including a flame vent and a spear\cite{moth_2019}.}
\item{Doomfist from Overwatch also sports enhanced arms. His need for arm replacement came due to the event known as the Omnic Crisis. His new arms enhance his power, especially in close combat\cite{moth_2019}.}
\end{itemize}
Again, whilst it is not said that a disabled character should be at a disadvantage in a game, the prevalence and dominance of this image perpetuates an unrealistic, non-real world view.

Towards the latter part of the survey, respondents were asked to what extent disabled characters were represented in games. The question was numerical with the lowest (one) being not enough and rising to five being enough. Across all the groups the response was very much towards "Not Enough". A small percentage fell into the higher segments, a response of four or five, though the highest response percentages were in the lower sections. When analysed against the decade that gamers started to play games, the percentages are consistent. The average for all groups for the lowest value was $48.29$ with the $2^{nd}$ being $35.84$.  The breakdown of the values can be seen in Figure \ref{fig:Year}.
 
\pgfplotstableread[row sep=\\,col sep=&]{
    interval & $1970's$ & $1980's$ & $1990's$ & $2000's$ & $2010's$  \\
    $Not Enough: 1$ & 50.0 & 78.57 & 55.17 & 57.69 & 0.0\\
    $2$ & 50.0 & 14.29 & 20.69 & 19.23 & 75.0 \\
    $3$ & 0.0 & 7.14 & 13.79 & 23.08 & 25\\
    $4$ & 0.0 & 0.0 & 3.45 & 0.0 & 0.0\\
    $Enough: 5$ & 0.0 & 0.0 & 6.90 & 0.0 & 0.0\\
    }\mydata 
\begin{center}
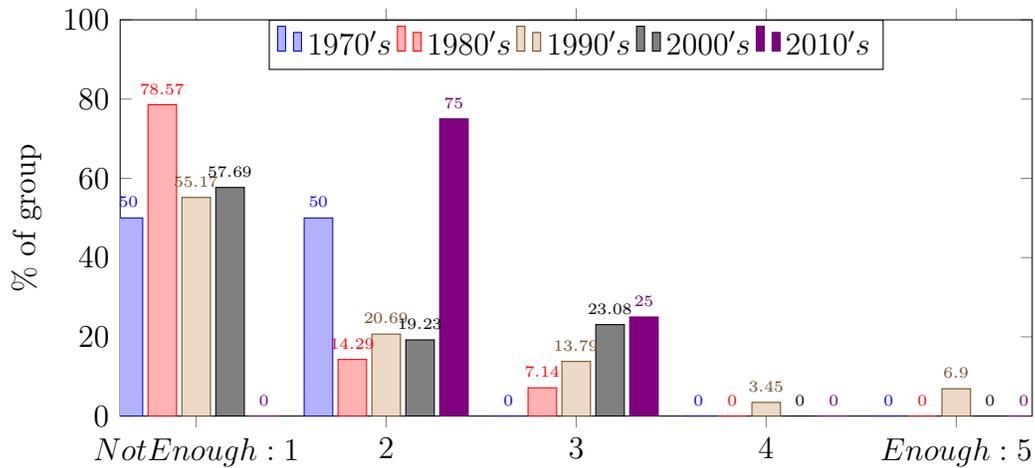
\begin{figure}[ht]
\begin{tikzpicture}
    \begin{axis}[
            ybar,
            bar width=.38cm,
            width=\textwidth,
            height=.5\textwidth,every node near coord/.append style={font=	                    \tiny},
            legend style={at={(0.5,1)},
            anchor=north,legend columns=-1},
            symbolic x coords={$Not Enough: 1$,$2$,$3$,$4$,$Enough: 5$},
            xtick=data,
            nodes near coords,
            nodes near coords align={vertical},
            ymin=0,ymax=100,
            ylabel={\% of group},
            every node near coord/.append style={font=\tiny},
        ]
        \addplot table[x=interval,y=$1970's$]{\mydata};
        \addplot table[x=interval,y=$1980's$]{\mydata};
        \addplot table[x=interval,y=$1990's$]{\mydata};
        \addplot table[x=interval,y=$2000's$]{\mydata};
        \addplot table[x=interval,y=$2010's$]{\mydata};
        \legend{$1970's$,$1980's$,$1990's$,$2000's$,$2010's$}
    \end{axis}    
\end{tikzpicture}
\caption{The Extent to Which Disabled Characters Are Deemed to Be Represented Across Decades Participants Started to Play.\label{fig:Year}}
\end{figure}
\end{center}
This demonstrates that there has been be a perceived change in the representation across the five decades, the predominant view that the representation of disabled characters is too low. Again, there is a correlation to other minorities such as females, however some research has shown that there is an increase in female playable characters since the $1990's$.

\section{Discussion}
\label{Sec:Discussion_Survey}
Many aspects of the survey supported those found in the analysis of the game trailers. The survey highlighted that very little has changed in the character portrayal aside from a perceived increase in the presence of women. It demonstrated that game players are aware of changes, but do not necessarily view these as always beneficial. The greater inclusion of women was viewed by individual respondents as producing more of a sexualised representation.

A high number of respondents cited a lack of disability representation in games. This supports the low numbers that were found in the trailer survey. The survey also highlighted that there appears to be a link between those that define themselves as having a disability and the playing of disabled characters. It is thought that this could be attributed to a disabled gamers striving to have their own disability represented, much in a similar way to both gender and ethnicity.
\section{Conclusion}
\label{Sec:Conclusion}
This research used both a content based analysis and a qualitative approach to construct the results. 108 trailers ranging from 2006 to 2016 were assessed for the inclusion and representation of disabled characters. Of the 108 trailers analysed, 20 of the trailers ($18.52$\%) were deemed to have a disabled character. Of this set, 15 were main characters with $10$ being playable. Opposed to many of the views held, of the $20$ trailers assessed the prominent disabled characters were protagonists rather than antagonists. 
Much of the discussion that has occurred throughout this piece of research is echoed in the study of both gender and ethnicity. Taking those both as a comparison, it is clear that although there have been steps forward in the inclusive nature of video games in the area of disability, there are areas were further examination needs to be carried out. These areas fall into two main categories:
\begin{itemize}
	\item Despite an increase in minority groups representation, characters with the disabilities are still very low. This is both a perceived view of video game players and how characters are portrayed by games studios.
	\item Due to the small amount of work in this area, to the authors knowledge there is no guidance for video game studios to work from. There is forming a kernel of a guidance to producing accessibility for video games\cite{noclip_2018}\cite{specialeffect2019}\cite{uncannyvivek2019vivek} but this is still in its infancy.
\end{itemize}

Additionally, the research adopted a qualitative approach through the use of a survey. 24 questions were used to gain insight into the view of representation. Overall, 75 responses were gained. This survey echoed many of the findings of the trailer analysis. 
A number of findings can be drawn:

\begin{itemize}
    \item Very little has perceived to have changed in respect to character representation across game genres, although there was discussion of an increase in the representation of women.
    \item The greater representation of women was not always positive as it can continue or increase the sexualised view of those characters.
    \item There appears to be a correlation between disabled gamers and the playing as disabled characters (this has to be taken against confusion regarding a question structure. See \ref{fig:disabledCharacter}).
    \item The decade a player started to play games had little bearing on their view of the quantity of disabled character representation. All groups responded highly as "not enough" when asked. 
    
\end{itemize}

\section{Future Work}
\label{Sec:Further}
The similarities between gender, ethnicity and disability were highlighted during this research. Further work to examine this would help to provide a more universal approach to the depiction of video game character minorities. In a similar strand, more work could be carried out working with the disabled community directly, to develop a set of "wants" to supply to the video games industry as a form of guidance.  This may include an investigation into the technical requirements of implementing a more sophisticated, nuanced customisation process.
Time constraints and the availability of data meant that 108 video game trailers were reviewed. A greater number could add further credence to the findings. Additionally the time period used was up until 2016. An expansion of this time frame could offer greater insight to how characters are depicted presently.






\bibliographystyle{elsarticle-num-names}
\bibliography{gamesV2.bib}


\end{document}